\documentclass[]{JHEP3}
\usepackage{psfrag,epsfig,epsf,citesort}
\newcommand{\lsim}{\buildrel < \over {\hspace{-.1em} {}_{\sim}} }
\newcommand{\gsim}{\buildrel > \over {\hspace{-.1em} {}_{\sim}} }
\newcommand{\tr}{{\mathrm{tr}}}

\newcommand{\be}{\begin{equation}}
\newcommand{\ee}{\end{equation}}
\newcommand{\bea}{\begin{eqnarray}}
\newcommand{\eea}{\end{eqnarray}}

\newcommand{\bear}{\begin{array}{l}}
\newcommand{\eear}{\end{array}}

\newcommand{\ie}{{\it i.e.}\ }
\newcommand{\cf}{{\it cf.}\ }
\newcommand{\eg}{{\it e.g.}\ }
\newcommand{\etal}{{\it et al.}\ }
\newcommand{\aka}{{\it a.k.a.}\ }
\newcommand{\etc}{{\it etc.}\ }
\newcommand{\viz}{{\it viz.}\ }

\def\eq#1{(\ref{#1})}
\def\eqs#1#2{(\ref{#1},\ref{#2})}
\def\sec#1{sec.\ \ref{#1}}
\def\fig#1{fig.\ \ref{#1}}

\def\TeV{\,\mathrm{TeV}}
\def\GeV{\,\mathrm{GeV}}
\newcommand{\A}{ {\cal A}}
\newcommand{\F}{ {\cal F}}
\newcommand{\LL}{ {\cal L}}

\newcommand{\ha}{ {\hat{a}}}
\newcommand{\bp}{ {\bar{p}}}
\newcommand{\bq}{ {\bar{q}}}
\newcommand{\bk}{ {\bar{k}}}
\newcommand{\ta}{ {\tilde{\alpha}}}
\newcommand{\tW}{ {\tilde{W}}}
\newcommand{\tB}{ {\tilde{B}}}
\newcommand{\tH}{ {\tilde{H}}}
\newcommand{\tv}{ {\tilde{v}}}
\newcommand{\tb}{ {\tilde\beta} }
\newcommand{\tl}{ {\tilde\lambda} }
\newcommand{\tg}{ {\tilde{g}}}

\title{Renormalizable extra-dimensional models}

\author{Tim R. Morris\\
Department of Physics, CERN, Theory Division, 1211 Geneva 23,
Switzerland, and School of Physics and Astronomy, University of
Southampton,
Highfield, Southampton SO17 1BJ, U.K.\\ E-mail:
\email{T.R.Morris@soton.ac.uk}}

\maketitle

\preprint{CERN-PH-TH/2004-120\\ SHEP 0420}

\abstract{Non-Abelian gauge theories may have continuum limits in
more than four dimensions, supported by non-trivial ultra-violet
fixed points. Moreover, such theories can be expected to be
accessible to Wilson's epsilon expansion. We investigate this
series for $SU(N)$ Yang-Mills, in particular for the fixed point
coupling and exponent $\nu$, up to four loops. From the
model-building point of view, such theories would be effectively
perturbatively renormalizable in the normal way. A particularly
attractive possibility is the construction of renormalizable
extra-dimensional models of the weak interactions, which have the
potential to address the full hierarchy problem. The simplest such
gauge-Higgs unification model is however ruled out by a
combination of theoretical and phenomenological constraints.}

\begin{document}
\section{Introduction}
\label{INT}

There has been considerable interest in the last few years in
constructing field theories of particle physics in more than four
dimensions. On the one hand, these ideas are inspired by similar
features in string theory, on the other hand, elegant new
approaches to the persistent mysteries of theoretical particle
physics are now possible \cite{Anton,Feruglio:2004zf}. These aim
variously to solve the famous hierarchy problem (\ie the fact that
the Higgs mass is quadratically sensitive to higher scales), the
little hierarchy problem (the fact that precision measurements
seem to favour a scenario with a low Higgs mass $\sim$ 115 GeV and
no new physics until $\sim$ 10 TeV \cite{Barbieri:1999tm}), the
pattern of fermion masses and mixing angles, and problems with
grand unified theories such as doublet-triplet splitting, proton
lifetime and mass relations. The new approaches to solving these
issues include theories with so-called universal extra dimensions
which e.g. result in calculable SUSY breaking and Higgs masses
\cite{Barbieri:2002uk}, gauge-Higgs unification in which the Higgs
field is a component of a higher dimensional gauge field whose
mass is thus protected by gauge invariance
\cite{Manton:1979kb,MasieroS3,AntoniadisBQ,HallNS,N1,BurdmanNomuraSUSY,GIQResidualGaugeSym,GIQ6D,S3},
Higgsless models where the symmetry is broken by the boundary
conditions \cite{Csaki}, grand unified models in which colour
triplet Higgs are given a Kaluza-Klein mass whilst the doublet
remains massless at the compactification scale \cite{kawaDTsplit}
(see also \cite{so10E6}), and approaches to the flavour problem
using the freedom to place fermions at different points (branes)
in the extra dimensions and to couple them non-locally through
other fields or Wilson lines \cite{EDnu,Larry,halln,ark2,N1,S3}.

However all of these new approaches suffer from a severe drawback:
quantum field theory as conventionally envisaged is not
renormalizable in more than four space-time dimensions. Thus these
theories must be viewed as effective field theories, only
applicable over a limited energy range. For energies much lower
than the defining scale $E$ in the theory (e.g. $E\sim1/R$ where
$R$ is the compactification scale) experimental constraints
require that it reduce to the Standard Model plus very small
corrections. On the other hand there is a maximum energy
$\Lambda$, usually taken to be the energy where some coupling
becomes strong and estimated by so-called naive dimensional
analysis (NDA) \cite{NDA}, above which something other than
quantum field theory must take over (for example some conjectural
string theory). This maximum energy depends on the couplings, with
larger $\Lambda/E$ for smaller couplings, but typically
$\Lambda\sim 10E $ to $100E$ is possible. With the ultraviolet
completion unknown, we can only bound the symmetry-preserving
higher dimensional vertices in the effective theory, under the
assumption that they naturally have order one coefficients at the
cutoff scale. This sets an irreducible limit on the expected
predictivity of these models controlled by $E/\Lambda$, again
dependent on the details, but typically authors aim for $\sim1\%$.

This limitation is less of a problem for theories that attempt to
describe the energy range between the GUT scale $10^{15}\GeV$ and
the Planck mass $10^{19}\GeV$, and for which the irreducible
uncertainty of $\sim1\%$ comes on top of already (better be) small
corrections to Standard Model quantities.

It is a much more drastic issue for higher dimensional theories
that directly address the hierarchy problem. These models tackle
the problem by introducing new physics at $E\sim1\TeV$ in such a
way that the Higgs mass becomes insensitive to higher scales. Here
attempts to bring such a model into agreement with the impressive
precision LEPI/II data can in turn force it into regions of
parameter space where the need for large ratios in couplings then
result by NDA in significant reductions in $\Lambda$, cutting off
its domain of applicability and thus severely limiting its
predictability (e.g. see \cite{S3,Rattazzi}). Since an as yet
unknown ultraviolet completion at some intermediate scale
$\Lambda$ is required, these models can actually only address the
little hierarchy problem, because any insensitivity of the Higgs
mass to scales higher than $\Lambda$ has to be explained in the
unknown theory. It is in any case a rather drastic step to address
the little hierarchy problem by abandoning quantum field theory
altogether at say $\sim100\TeV$ for a largely conjectural
ultraviolet non-field theory.

It might be possible however, to rest certain restricted classes
of such models on a more secure foundation. In the 1970s it seems
to have been commonly understood that non-Abelian Yang-Mills
theory may exist as a non-perturbative quantum field theory, \aka
continuum limit, in more than four space-time dimensions
\cite{Peskin}. With few exceptions since then
\cite{Gies,Creutz,Kawai,Nishimura,Ejiri,Hashimoto,Dimi,Farakos}
this possibility seems to have been ignored or forgotten.
Independently in string theory, probably related supersymmetric
versions have been discovered \cite{Seiberg}.\footnote{A
Kaluza-Klein treatment appears in ref. \cite{Dienes}.}

We briefly review the little evidence that has collected, for and
against such theories, in \sec{RED}. We also investigate the
series for the fixed point coupling $\ta_*$, and the critical
exponent $\nu$, using the $\epsilon$ expansion up to four loops.
Together, the evidence suggests that these fixed points exist in
$D=5$ dimensions, possibly in $D=6$ dimensions and probably not
for $D\ge7$ dimensions.

At any rate such theories formally do exist within an expansion in
$\epsilon$ of a $4+\epsilon$ dimensional theory. As we review in
\sec{RED}, this means that quantum corrections can be computed
perturbatively and renormalized in {\sl four} dimensions. From the
model building point of view, the theory is perturbatively
renormalizable; one only needs to bear in mind that at high
energies the appropriate dimensionless ratio $\ta$, involving the
Yang-Mills coupling $\alpha$, runs to a computable finite value
rather than zero.

For the continuum limit to exist within the $\epsilon$ expansion
when other fields are included, one requires that in four
dimensions, all couplings are relevant (effectively masses and
couplings of positive mass dimension) or marginally relevant
(equivalently asymptotically free). We outline the restrictions
this imposes in \sec{MBC}. We then extend this understanding of
renormalizability to models with branes, and discuss the
constraints that arise, concentrating on the bosonic sector, as
well as making some further comments on the applicability of this
idea to Grand Unified Theories.

A particularly attractive application is to extra dimensional
models of the weak interactions: in this case the insensitivity of
the Higgs mass to scales up to $\Lambda$, become solutions to the
hierarchy problem, since by the definition of renormalizability,
$\Lambda$ can be taken to infinity. We explore this in \sec{MBC}
and \sec{GHU}. We are led to consider models of the weak
interactions based on the Hosotani mechanism
\cite{Hosotani,Kubo:2001zc,Haba:2002py}.

We show however in \sec{GHU}, where the Hosotani mechanism, and
the fermionic sector are considered in detail, that the simplest
such model, based on a bulk $SU(3)$ gauge theory, is ruled out. In
particular we show that in order to get the right Yukawa
interactions, we are forced to add sufficiently many bulk matter
fields that the $SU(3)$ Yang-Mills is no longer asymptotically
free in four dimensions and thus has no continuum limit within the
$\epsilon$ expansion of $4+\epsilon$ dimensions. As we note in our
conclusions (\sec{CON}) this only means that one should try to
construct more involved models, for example with a larger gauge
group, and in addition introduce the bulk fields more
economically.

We start in \sec{RED} with a short explanation of the $\epsilon$
expansion, comparing to its use in its more traditional context.
We then review the evidence such as it is for higher dimensional
fixed points. After this we use the known four-loop
four-dimensional beta function \cite{upto3,Vermaseren} to
investigate the $\epsilon$ expansion for these fixed points. (It
seems that this is the first time the early two-loop investigation
\cite{Peskin} has been extended.)

\section{Renormalizability in extra dimensions}
\label{RED}

Scalar field theory in three dimensions has a Wilson-Fisher fixed
point \cite{Wil}. In the case of an  $N$-component scalar field
theory with $O(N)$ invariance, tuning the couplings so that the
theory lies close to this fixed point results in universal
behaviour in a universality class determined only by $N$ (see \eg
ref. \cite{ZJ}). The `distance' from the fixed point sets a
mass-scale $m$ (the inverse correlation length) which is in fact
the only parameter left in the universal physics.\footnote{There
is one relevant coupling, which can be parametrized by $m$, and
there are no marginal couplings.} Exactly at the fixed point, one
obtains an interacting conformal field theory. The properties of
the conformal field theory determine everything that is universal
about this fixed point, for example the scaling dimensions of its
operators determine the various critical exponents $\nu, \omega,
\eta$ \etc, pure numbers that are amongst the simplest quantities
to calculate and/or measure, while the effective potential for
$\phi$ yields the universal equation of state. Viewed in this way,
this conformal field theory and its leading relevant deformation
constitute a concrete example of non-trivial quantum field theory,
\ie a non-trivial continuum limit, about a fixed point other than
the Gaussian fixed point.

As we will see, in a closely analogous way, non-trivial continuum
limits may well exist for Yang-Mills theory in higher than four
dimensions. In order to gain a complete appreciation of these
phenomenona one needs to use the language of the Wilsonian
renormalization group \cite{Wil}. For clarity we remind the reader
of the basic elements. One works within an infinite dimensional
space of bare actions that include all possible local interactions
allowed by the symmetries of the theory. In this space, there is
the so-called critical manifold, which consists of all bare
actions yielding a given conformally invariant continuum limit
(for example the $O(N)$ conformal field theories above or the
non-perturbative gauge theory cases we are about to discuss). Any
point on this manifold -- \ie any such bare action -- flows under
a given Wilsonian renormalization group towards its fixed point;
local to the fixed point, the critical manifold is spanned by an
infinite\footnote{this is strictly an assumption on our part for
the gauge theory cases of interest} set of irrelevant operators.
The other directions emanating out of the critical manifold at the
fixed point, are spanned by relevant and marginally relevant
perturbations (couplings with positive and vanishing scaling
dimensions respectively, but where the latter grow as we move to
lower energies, \ie are asymptotically free). Next, we choose a
bare action with sufficiently many operators of the right type to
intersect the critical surface, for some choices of couplings.
Note that in general we have no way of knowing what type of bare
action will do: for a non-perturbative fixed point we cannot rely
on na\"\i ve dimensional analysis to classify which operators are
(marginally) relevant. Instead we simply need to search the full
space.

Now in the bare action, we shift a coupling a little bit away from
the critical manifold. The trajectory of the Wilsonian
renormalization group will to begin with, move towards the fixed
point, but then shoot away along one of the relevant directions
towards the so-called high temperature fixed point which
represents a theory with only infinite mass scales.

To obtain the continuum limit, and thus a finite mass scale, one
must then tune the bare action towards the critical manifold and
at the same time, reexpress physical quantities in renormalized
terms appropriate for the diverging correlation length.

To {\sl confirm} the existence of such a continuum limit, one
really has no choice but to follow through the above procedure for
example within the framework of a lattice gauge theory computation
in higher than four dimensions.

Unfortunately, to date all such investigations have been limited,
with ambiguous conclusions. Thus in ref. \cite{Gies}, an exact
renormalization group treatment was used, which taken at face
value implies that the non-trivial fixed point exists for $SU(N)$
Yang-Mills in $D=5$ dimensions for $N\le5$. The author computed
the critical dimension $D_{cr}$ above which the fixed point
disappears, finding $5<D_{cr}<6$ in all these cases. Actually, in
ref. \cite{Gies} $N=4$ is not reported but the results for
$N=2,3,5$ do suggest that the critical dimension $D_{cr}$ behaves
smoothly with $N$ as expected. The problem with such a treatment
is primarily that there is no control of uncertainties due to
truncation of the flow equations. There are a small number of
lattice studies \cite{Creutz,Kawai,Nishimura,Ejiri}. These are
subject of course to limitations from systematic and statistical
errors, which rapidly become more severe with increasing dimension
(and increasing $N$). The general conclusion seems to be that a
non-trivial continuum limit does not exist in $D>4$ dimensions for
the simple Wilson plaquette action (at least for $N=2$ and
$D=5,6$). However as we have already emphasised, there is no
reason to expect that this simplest action is the correct one in
this case.\footnote{See also ref. \cite{Gies}. This is independent
of a possible multicriticality of the fixed point however.} Ref.
\cite{Ejiri} displayed evidence for a non-trivial continuum limit
even with the simple Wilson plaquette action, providing the extra
dimensions are small enough, but it is reasonable to assume that
this arises from lattice artifacts \cite{Farakos,Gies}. Ref.
\cite{Kawai} reported on the addition of an adjoint representation
Wilson plaquette but the results were inconclusive. Only in ref.
\cite{Nishimura} were negative conclusions drawn from more general
actions than the simple Wilson plaquette action, but this was in
the special case of $D=6$ dimensions with large $N$ (=27 and 64)
after reduction to a twisted Eguchi-Kawai model.

If these non-trivial fixed points exist, one of the most powerful
ways of deriving the physics resulting from these fixed point
behaviours is by computing perturbative quantum field theory in
four dimensions and using the so-called $\epsilon$ expansion
\cite{ZJ}.

The idea is as follows. In four dimensions we know that the
perturbative Yang-Mills theory is based around the Gaussian fixed
point, and parametrized by a small marginally relevant coupling
$g$ (just as perturbative scalar field theory is parametrized by a
small marginally irrelevant coupling $\lambda$ and a relevant mass
term). All the independent higher point interactions (field
strength cubed \etc, in the scalar case $\phi^6$ \etc) are
irrelevant; their effect just amounts to a finite renormalization
of the low energy couplings. Because the theory is free at the
associated fixed point, the scaling dimensions of the irrelevant
operators are equal to their engineering dimensions, and thus are
6 or greater. Now, providing that the scaling dimensions of these
operators are continuous functions of the dimension $D$, it must
be that they are still irrelevant in $D=4+\epsilon$ dimensions,
for small enough $\epsilon$. Thus for small enough $\epsilon$, the
usual bare action and renormalized effective Lagrangian must still
yield the right description.

In the scalar case, this results in the classic argument for
understanding the existence of the Wilson-Fisher fixed point: the
beta function takes the form
\be
\label{lamb}
\mu \partial_\mu\tl = \epsilon\tl + \beta(\tl),
\ee
where the first term is classical and arises simply because we use
the Wilsonian coupling, namely the dimensionless combination $\tl
= \lambda\mu^\epsilon$, while the second --quantum-- term can
coincide with the beta function in four dimensions (as we will see
explicitly later). Since below four dimensions the classical term
is negative while the quantum term starts at higher powers than
$\tl$ and is positive (reflecting triviality of four dimensional
scalar field theory), we get a zero at some point $\tl=\tl_*$, at
least for sufficiently small $\epsilon$, corresponding to an
infrared attractive fixed point for $\tl$.

In the gauge theory case, we have qualitatively the same situation
but with the signs reversed. The quantum term can be chosen to be
the four-dimensional beta function and is now negative
(corresponding to asymptotic freedom) and this can be balanced by
going above four dimensions, where the classical term is positive
(corresponding to a negative dimension coupling $g$). Again, we
get a fixed point $\tg = \tg_*$, for sufficiently small
$\epsilon$, but this time it corresponds to the desired
ultraviolet attractive fixed point. Indeed we can now compute this
fixed point and its universal consequences by the $\epsilon$
expansion, \ie simply by solving for $\tg_*$ order by order in
$\epsilon$.

In fact for the gauge theory, $\epsilon$ expansion of non-linear
sigma models is even more closely analogous
\cite{ZJ,Wegner,Brezin}. $O(N)$ invariant non-linear sigma models
in two dimensions are also renormalizable and asymptotically free.
By the same logic, we expect that in $D=2+\epsilon$ dimensions,
there is still a continuum limit for these sigma models but now
based around a non-trivial ultraviolet attractive fixed point.
This fixed point indeed exists and is nothing but the infrared
description of the $O(N)$ Wilson-Fisher fixed points already
discussed (as may be confirmed for example by noting that the
large $N$ expansion of this fixed point coincides with the large
$N$ expansion of the one obtained in $D$ dimensional scalar field
theory) \cite{ZJ}.

(There is some controversy over the reasons for the poor accuracy
in practice of the $\epsilon$ expansion for these non-linear sigma
models \cite{Wegner,Brezin}; some crucial non-perturbative effect
is perhaps missing. At any rate the qualitative conclusion is
correct, namely that ultra-violet fixed points exist above the
upper-critical dimension where these theories cease to be
perturbatively renormalizable. It remains to be understood whether
the issues of accuracy have any bearing on the case of Yang-Mills
we discuss here.)

With respect to the use of the $\epsilon$ expansion in all these
examples, a word of caution is in order about taking the
description too literally. Although the $\epsilon$ expansion can
furnish impressively accurate answers for universal quantities
\cite{ZJ} (the renormalized quantities, which are all we are
interested in), of course the four dimensional field theory has
very different properties in general from the $D\ne4$ theory
(which latter needs to be understood from a Wilsonian
renormalization group perspective, as already emphasised). Thus,
for example for the $O(N)$ scalar field theory in three
dimensions, perturbation theory about the Gaussian fixed point
generically includes the now marginal $\phi^6$ interaction. More
importantly, perturbative three dimensional $\phi^4$ field theory
is superrenormalizable. Its ultraviolet divergences do not give
rise to the four dimensional $\beta$ function in \eq{lamb}. Indeed
$\lambda$ suffers no ultraviolet divergences at all; only $m^2$
receives divergent corrections at one and two loops. Instead, its
massless limit is plagued with infrared divergences, a signal that
we are working about the wrong fixed point. In $4+\epsilon$
dimensions, with $-1<\epsilon<0$, even for arbitrarily small
$\epsilon$, $\phi^4$ scalar field theory is still
superrenormalizable and its massless limit suffers infrared
divergences, first arising at $\sim-2/\epsilon$ loops (as follows
simply from power counting) \cite{ZJ}. In gauge theory in five
dimensions we cannot {\it a priori} exclude any of the full
infinity of higher dimensional gauge invariant operators from a
valid bare action. The theory in five dimensions is perturbatively
non-renormalizable; it is plagued with infinitely many new types
of ultraviolet divergence. In particular, its ultraviolet
divergences do not give rise to the four dimensional $\beta$
function in \eq{be}. Even in $4+\epsilon$ dimensions, with
arbitrarily small $\epsilon>0$, Yang-Mills theory has
non-renormalizable divergences that appear first at
$\sim2/\epsilon$ loops (again simply by power counting).

However, the $\epsilon$ expansion allows us to exchange the limits
($\epsilon\to0$ and $g\to0$) and follow perturbatively the
relevant fixed point, and thus in this way access the required
continuum limit (\ie the renormalized quantities) directly.

We now turn to the details. We write the gauge field in $D$
dimensions as $\A_M=\A^a_MT^a$, where the $T^a$ are generators of
some simple gauge group $G$, with standard normalisation $\tr\,
T^aT^b= \delta^{ab}/2$. (We use a metric of signature
$+{}-{}-{}\cdots$.) Writing the covariant derivative as $\nabla_M
= \partial_M - i \A_M$, the field strength is $\F_{MN} =
i[\nabla_M,\nabla_N]$. In $D=4$ dimensions the Lagrangian density
takes the usual form
\be
\label{bulk}
\LL = -{1\over 2g^2}\,\tr\,\F^{MN}\F_{MN}.
\ee
As we have already stated, we expect that the situation is
continuous in $\epsilon$ and thus in $D=4+\epsilon$ dimensions,
\eq{bulk} is still the right description for small enough
$\epsilon$. At this point it is helpful to recall a textbook
argument. In dimensional regularisation and minimal subtraction,
we replace the coupling in \eq{bulk} by $g_0$, and use the
dimensionless renormalized $\tg = \mu^{\epsilon/2} g$ to derive
\be
\label{g0}
g_0 = \mu^{-\epsilon/2}\left(\tg+\sum_1^\infty
{Z_n(\tg)\over\epsilon^n}\right)\ ,
\ee
where the $Z_n(\tg) = O(\tg^{2n+1})$ are the usual power series in
$\tg$ with numerical coefficients chosen to cancel all the poles
that arise order by order in perturbation theory as
$\epsilon\to0$. Now we differentiate \eq{g0} with respect to
$\mu$. Since the theory is finite when expressed in terms of
$\tg$, and using the fact that $g_0$ does not depend on $\mu$, we
can invert the result to obtain
\be
\label{be}
\mu\partial_\mu\tg = {\epsilon\over2} \tg + \beta(\tg),
\ee
where $\beta(\tg) = Z_1(\tg)/2-\tg Z'_1(\tg)/2$, and all the $Z_i$
with $i>1$ are determined in terms of $Z_1$ by the requirement
that all the $1/\epsilon$ poles cancel in \eq{be}.

As promised, we see that in a little more than four dimensions
($\epsilon>0$) we can balance the first positive term against the
negative power series $\beta(\tg)$ to get a fixed point $\tg_*$,
whose properties we can now compute order by order in $\epsilon$.

For what ensues, it is especially important to note that even
though \eq{be} expresses the flow of $\tg$ in more than four
dimensions, $\beta(\tg)$ is {\sl precisely the usual four
dimensional beta function}. Thus to deduce the properties of the
model in greater than four dimensions, we only need renormalize
the model in the normal way in four dimensions and then use
\eq{be} and similar equations as a formal device to analytically
continue to $\epsilon>0$.

As we will now show, the $\epsilon$ expansion for these
ultraviolet fixed points is in fact surprisingly well behaved. We
can illustrate this using the known four-loop $\beta$ function
\cite{upto3,Vermaseren} for $SU(N)$ Yang-Mills.

At this point we note a convenient point of detail: the
four-dimensional $\beta$ function is the same in the MS and
${\overline {\rm MS}}$ schemes \cite{Vermaseren}. This can be
understood from the definition of ${\overline {\rm MS}}$ namely
that, instead of subtracting via counterterms, powers of
$2\tg^2/(4\pi)^2$ times poles in $\epsilon$, one substracts powers
of
$$
{2\tg^2\over(4\pi)^2}\{ 1 + \ln(4\pi)\epsilon - \gamma_E\epsilon\}
$$
($\gamma_E$ being Euler's constant), the corrections arising from
$$
\Omega_D = {2\over(4\pi)^{D/2}\Gamma(D/2)},
$$
the $D$ dimensional solid angle divided by $(2\pi)^D$ that
accompanies each loop integral. Therefore for the $Z_n$ in
\eq{g0}, ${\overline {\rm MS}}$ amounts to an $\epsilon$ dependent
rescaling of $\tg$. When working at finite $\epsilon$, the results
can be expected to be better behaved if we instead absorb the full
dependence coming from $\Omega_D$ \cite{ZJ}. This amounts to
replacing $2\tg^2/(4\pi)^2$ in the MS scheme, with
$\tg^2\Omega_D$. Clearly this modified MS also yields the same
four-dimensional $\beta(\tg)$. Finally for convenience in the
ensuing analysis, we also absorb a factor of $N$ and write
\be
\label{ta}
\ta :=  N\Omega_D\,\tg^2 = N\Omega_D\,\mu^\epsilon g^2
\ee

In terms of $\ta$, \eq{be} becomes
\be
\label{beta}
\mu\partial_\mu\ta = \epsilon \ta + \beta(\ta),
\ee
where
\be
\beta(\ta) =
-\beta_0\ta^2-\beta_1\ta^3-\beta_2\ta^4-\beta_3\ta^5+O(\ta^6);
\ee
these $\beta_n$ are the coefficients quoted in ref.
\cite{Vermaseren} but divided by $2^nN^{n+1}$.

We compute $\ta_*$, although this is not universal, but we also
compute the index $\nu$ via \cite{Peskin}
\be
\label{nu}
{1\over\nu} := - {d\beta\over d\tg}(\tg_*) = - {d\beta\over
d\ta}(\ta_*),
\ee
which is the universal power relating the analogous QCD-like scale
$\Lambda_{YM}$ to the distance from the fixed point at any high
scale $\mu$ where the coupling is $\ta(\mu)$ is sufficiently close
to $\ta_*$:
\be
\Lambda_{YM} \propto \mu|\ta-\ta_*|^\nu.
\ee
Solving \eq{beta} for $\ta_*$,
\be
\label{fpta}
\ta_* = {\epsilon\over\beta_0}-{\beta_1\epsilon^2\over\beta_0^3}
+\left(2{\beta_1^2\over\beta_0^5}-{\beta_2\over\beta_0^4}\right)\epsilon^3
+\left(5{\beta_2\beta_1\over\beta_0^6}-{\beta_3\over\beta_0^5}-5{\beta_1^3\over\beta_0^7}\right)\epsilon^4
+O(\epsilon^5),
\ee
and substituting the values for $\beta_n$, we find a surprise.
Whereas the coefficients $\beta_n$ show the expected dramatic
increase in $n$, consistent with \eq{beta} being asymptotic (even
at $N=\infty$):
\be
\beta_0 = 3.667,\quad\beta_1= 5.667,\quad\beta_2=13.23,\quad\beta_3=39.43+51.22/N^2,
\ee
the series for $\ta_*$
\be
\label{fptan}
\ta_* = .2727\epsilon -.1150\epsilon^2+.02372\epsilon^3
-(.007395+.07729/N^2)\epsilon^4,
\ee
is considerably better behaved, and at $N=\infty$ it is not at
first clear that this series is asymptotic. Substituting \eq{fpta}
in \eq{nu}, we obtain
\be
\label{fpnun}
{1\over\nu}=\epsilon+.4215\epsilon^2+.1813\epsilon^3+(.1242+.8502/N^2)\epsilon^4.
\ee
Again, at $N=\infty$, this series looks very well behaved.

We start by investigating this limit in more detail. Writing the
coefficients of $\epsilon^n$ in \eq{fptan} and \eq{fpnun} as $a_n$
and $\nu_n$ respectively, successive ratios $|a_n/a_{n+1}|$  give
2.37,\,4.84,\,3.21 for $n=1,2,3$. And similarly successive ratios
$|\nu_n/\nu_{n+1}|$ give 2.37,\,2.32,\,1.45. If we did not know
better, we might even be tempted to conclude that these ratios
have a limit $r>0$, corresponding to a finite radius of
convergence $r$. However, we see from \eq{fptan}, that $\ta_*$ has
negative values for all $\epsilon<0$, as expected, since the four
dimensional beta function analytically continued to $\alpha<0$
behaves like that of a trivial theory and thus allows us to
balance the two terms in \eq{beta} and find a fixed point in $D<4$
dimensions. Since the theory is non-perturbatively sick for
$\alpha<0$ \cite{Dyson}, the above series in $\epsilon$ must have
zero radius of convergence.

If the series are already displaying their limiting asymptotic
behaviour (\viz $a_n\sim cs^n \Gamma(n+\zeta)/R^n$, for $s=-1$ and
some constants $c$, $\zeta$ and $R$, and similarly for $\nu_n$,
with $s=1$)\footnote{equivalently $a_n\sim c(-)^n n^\zeta n!/R^n$}
then we can estimate them by taking the sum to the point where the
terms stop getting smaller (including the smallest term) and using
the next term to estimate the error. In this case we ought to see
$|a_n/a_{n+1}|\sim R/(n+\zeta)$, and thus
$\lim_{n\to\infty}n|a_n/a_{n+1}|=R$. Forming this combination for
$n=1,2,3$, we get for the $a_n$s: 2.37,\,9.69,\,9.62, and for the
$\nu_n$s: 2.37,\,4.65,\,4.38.

We can already draw a number of important conclusions. Firstly,
there is clear evidence that the series do approximate their
limiting behaviour already after the first term. Secondly, the
fact that the last two combinations are so close to each other,
for both series, indicates that the corresponding $\zeta$ must be
small in these cases; of course the series are not long enough to
reliably estimate it however. Thirdly, we expect the $\sim R^{\rm
th}$ coefficient to be the smallest one, thus for $1/\nu$ in
\eq{fpnun}, we expect the next (5-loop) term to have a larger
coefficient than the four-loop term, while for $\ta_*$, we expect
to have to go to 10 loops in \eq{fptan} to see the coefficients
start growing with $n$.

Fourthly, we can use the series to estimate $\nu$ and $\ta_*$ for
positive integer $\epsilon$, as follows. For $\nu$ we simply
estimate it as described above, from the asymptotic series for
$1/\nu$; for $\epsilon=1$ we have a special case in that we are
missing the 5-loop term that would provide an estimate of the
error. This is why there is no error for the appropriate entry in
table \ref{table:n}.

For $\ta_*$ at $\epsilon=1$ we have an alternating series in which
we expect the terms to keep getting smaller until the $(R-1)^{\rm
th}$ term with the slightly larger $R^{\rm th}$ term providing the
error, where $R\approx10$. Clearly in this case $\ta_*$ should lie
between the sum of the first three and the sum of the first four
terms.\footnote{If the coefficients kept getting smaller this
would be a theorem.} This gives us the $D=5$, $N=\infty$ estimate
in table \ref{table:a}. In general the point where the terms stop
getting smaller can be expected to be at $n=R/\epsilon$. Comparing
with \eq{fptan}, we see that for integer $\epsilon>1$ we should
use the above method for summing asymptotic series, except that
for $\epsilon=2$ and 3, we are missing the term that would provide
the error. This is why we do not display an error when $D=6$ and
$N=\infty$.

Finally, at finite $N$, the series are clearly asymptotic, the
$1/N^2$ terms in \eqs{fptan}{fpnun} (the result of the first
divergent non-planar contribution - which appears at four loops
\cite{Vermaseren}) having a much larger coefficient (and of the
same sign). Without further information (see below), we can
estimate the sums only in the cases where this last term is not
the smallest term. We find that the last term is smallest in
\eq{fpnun} only when $D=5$ and $N\ge4$, and in \eq{fptan} when
$D=5$ and $N\ge3$, and when $D=6$ and $N\ge5$. This is why we the
leave the corresponding entries blank in the tables.

\TABLE{

\begin{tabular}{|c|c|c|c|c|c|}
\hline
 $D\backslash N$ &   2    &    3    &    4    & 5 & $\infty$ \\
\hline
   5   & .18(3) &        &        &   & .178(4)  \\
\hline
   6   & .28(43)& .28(26) & .28(20) &  & .157  \\
\hline
\end{tabular}

\label{table:a}
\caption{Estimates for $\ta_*$ for varying $N$ and $D$.} }

In addition, we find that for $1/\nu$, the terms in the series
only increase once $D\ge7$ (for any $N$). For $\ta_*$ when $D=7$,
the error is larger than the estimate for $2\le N \le 12$, after
which the last term is the smallest so that we cannot provide an
estimate. At $N=\infty$, we saw above that we cannot estimate the
error, and the series sums to $\ta_*=-0.175$, which cannot make
sense physically. For $D=8$, the error is much larger than the
estimate for $\ta_*$ for all $N$. For $D\ge9$, the terms only
increase in magnitude with $n$. We take all this as evidence from
the $\epsilon$ expansion that the fixed points do not exist in
$D\ge7$.

\TABLE{

\begin{tabular}{|c|c|c|c|c|}
\hline
 $D\backslash N$ &    2    &    3    &    4    & $\infty$ \\
\hline
   5   & .62(13) & .62(9)  &        & .579  \\
\hline
   6   & .19(20) & .19(13) & .19(11) &  .19(8)  \\
\hline
\end{tabular}

\label{table:n}
\caption{Estimates for $\nu$ for varying $N$ and $D$.} }

In $D=6$ the evidence is somewhat marginal. At $N=2$ the error is
larger than the estimate, both for $\nu$ and $\ta_*$. Perhaps this
indicates that the fixed point does not exist. For $\nu$ the
errors are large but gradually decreasing for $N\ge3$ and $N$
increasing, but even at $N=\infty$ the error is 40\%. For $\ta_*$
the errors remain large where they can be estimated at all.

On the other hand, for $D=5$ we appear to have clear evidence for
a fixed point: the errors are small where they can be estimated.
There are missing entries only because the series are too short in
these cases. Given the other values it is reasonable to assume
that $\ta_*\sim.18$ and $\nu\sim.6$ for all $N\ge2$.

These conclusions are in broad agreement with the earlier
alternative approaches already discussed
\cite{Gies,Creutz,Kawai,Nishimura,Ejiri}, given the limitations
with all these studies.

We have only performed the simplest estimates. It should be
possible to do much better. In the lower dimensional scalar field
theories, powerful techniques have been developed to cope with
their divergent $\epsilon$ expansions.  One first transforms the
series to the Borel plane. Relying on no other knowledge the
series can be resummed by Pad\'e approximants. However, if one
assumes the $\epsilon$ expansion inherits the large order
behaviour of $\lambda\phi^4$ theory in three dimensions one knows
about the closest singularity in the Borel plane (from studying
Lipatov instantons) \cite{ZJ}. By conformal mapping the Borel
plane, this information can be taken into account. The results are
impressive: for better known quantities \eg $\tl_*$ and $\nu$ (and
using also information from the exactly soluble two-dimensional
Ising model) the theoretical error has been reduced to a few per
mille \cite{ZJ}.

In the gauge theory case, one can also transform the series to the
Borel plane and use Pad\'e approximants. However, one also has
similar knowledge about the large order behaviour of perturbation
theory coming from the closest singularities in the Borel plane,
in this case due to infrared renormalons and instantons, which can
also be taken into account.

\section{Renormalizability with matter and branes}
\label{MBC}

We pursue theories which have non-perturbative continuum limits in
more than four space-time dimensions, in the sense of having
ultraviolet fixed points within the $\epsilon$ expansion. We have
seen that this requires that when $\epsilon=0$, the model must be
renormalizable and in addition all couplings in the bulk are
relevant (effectively masses and couplings of positive mass
dimension) or marginally relevant (equivalently asymptotically
free).\footnote{Note that these are the same conditions that are
believed to be required to ensure the non-perturbative existence
of the quantum field theory in four dimensions.}

Thus we can readily incorporate non-Abelian gauge fields in the
bulk of the higher dimensions.  We can then add fermions to the
bulk of the higher dimensions, providing there are not too many or
that their representations are too large, such as would destroy
the coupling's asymptotic freedom. It is harder to include scalars
in the bulk (because the scalar self-coupling must also be
asymptotically free), but possible for a careful choice of
couplings. On the other hand, we cannot at the same time as
keeping asymptotic freedom, use the scalars to give a mass to all
the gauge fields via the Higgs mechanism \cite{ColemanGross}.
Inclusion of both fermions and scalars allows asymptotic freedom
only for a fine tuning of the Yukawa couplings which is natural
only in supersymmetric theories \cite{ColemanGross}. Finally,
Abelian gauge fields can be included only on four dimensional
branes if at all.

As is already clear, these models are highly restricted. We will
however meet yet other constraints renormalizability puts on the
couplings. We will concentrate on the case of five dimensional
theories, \ie where one should set $\epsilon=1$ eventually. This
is because, as we have seen, the evidence for non-trivial fixed
points is strongest in five dimensions. However, most of our
comments obviously extend to other dimensions. Our conventions are
to write five dimensional indices in Latin capitals with
components $M=0,1,2,3,5$; coordinates we will write as $x^M =
(x^\mu,y)$, while we write momenta $p_M = (\bp_\mu,p_5)$, the bar
being an extra flag for four-dimensional parts. We write the
Lagrangian density of the non-Abelian gauge field as in eqn.
\eq{bulk}.

We do not consider gravity since we do not know how to describe it
as a renormalizable theory. We thus take the conservative view
that there is a flat classical gravitational background, to be set
and not questioned. Phenomenologically therefore, these models
have to have the fifth dimension compactified, which we can take
to be a circle radius $R$: $y$ is identified with $y+2\pi R$.
Although the models we consider, are valid as quantum field
theories up to infinite energies, in practice we have to set the
cutoff at the natural five-dimensional Planck mass since quantum
gravitational effects cannot be ignored above this. (This means
that $\Lambda\sim a^{1/3}\times 10^{13}\GeV$, where we have taken
the radius of compactification to be $1/R = a$~TeV
\cite{Arkani-Hamed}.)

At first sight it is attractive when computing such a model in the
$\epsilon$ expansion, to regard the extra $\epsilon$ dimensions as
the ones that are compactified. This is not consistent however. In
the $\epsilon$ expansion the extra dimensions only enter by the
change in the scaling of the classical coupling constant. The
calculation is thus not sensitive to topological or geometric
features of the $\epsilon$ extra dimensions. This is reasonable:
the approach to the fixed point ${\tilde g}_*$, is a property of
the theory in the far ultraviolet, \ie at very small length
scales, where space-time should look flat. Where a more precise
matching is required between renormalized couplings above and
below a compactification scale, the result from $\epsilon$
expansion with one of the {\sl original} dimensions compactified,
should clearly be used to compute threshold corrections.

If parts of the five dimensional gauge fields $\A_M$ are to
represent some Standard Model gauge fields, then we need to go
beyond trivial compactifications.  (We do not observe the adjoint
massless scalars $\A_5$ that would arise from such compactified
Yang-Mills fields, and we cannot use extra-dimensional Higgs
fields to give masses to all parts of $\A_5$.) There are two
overlapping possibilities that have been suggested in the
literature, the extra dimensional space can be an orbifold or more
generally can have a boundary on which non-trivial boundary
conditions are imposed \cite{Csaki}.

Either way, once we take the model seriously as describing a
renormalizable continuum limit,  we have to introduce boundaries,
\aka branes, and boundary couplings which we cannot ignore, as we
now explain.

In the general case \cite{Csaki}, boundary conditions on the gauge
fields are imposed that explicitly break the gauge group $G$ down
to some smaller (\eg Standard Model) group $H$. Of course an
explicitly broken gauge theory is not renormalizable. This is the
only problem, and is not in fact a problem {\it per se}, for the
little hierarchy solution envisaged by these authors, as can be
understood by viewing the boundary conditions as imposed by
boundary Higgs fields in the non-linear sigma model limit
\cite{Bhiggs}. The effective cutoff for the sigma model can be
identified with the effective cutoff the theory already
necessarily has. However these problems do rule out general
breaking by boundary conditions as an option here, because the
non-linear sigma model is not renormalizable.

The orbifold alternative consists in dividing out the extra
dimension by a discrete symmetry which is not freely acting. For
illustration we consider only the simplest case of $Z_2$ parity $y
\mapsto -y$.\footnote{Most of our comments apply equally well to
more involved cases, the most general case in this situation being
$Z_2\times Z'_2$.} We are interested in the case where the parity
does not commute with the gauge group (otherwise all of $\A_5$
will be odd, thus gain Kaluza-Klein (KK) masses 
of order the compactification scale, 
and just return us at low energies to a four dimensional world of
Yang-Mills with gauge group $G$). The gauge fields satisfy
$\A_M(x,y) = P\A_M(x,-y)P$. (It is to be understood that $P$ also
maps any 5th Lorentz component to minus itself.) Providing $P$
generates an automorphism of the Lie algebra of $G$, such an
action is consistent, but the result is that the gauge symmetry is
typically restricted to a smaller semi-simple group $H$ at the
orbifold points $y=0,\pi R$. Indeed, we can always choose a basis
for the Lie algebra so that the action of $P$ is diagonal. The
even generators, $PT^{\ha}P=T^{\ha}$, generate the subgroup $H$,
whilst all the other generators $T^{a'}$ are odd. Although from
the low energy four-dimensional point of view, the gauge group $G$
appears to have been broken to $H$, this is not really true. Gauge
transformations $\delta \A_M = \nabla_M\Omega$ are only restricted
in the sense that $\Omega$ must also satisfy $\Omega(x,y) =
P\Omega(x,-y)P$, and thus the components $\Omega^{a'}$ vanish at
the boundaries $y=0,\pi R$.

As noted in ref. \cite{Georgi}, the orbifolding of the 5th
dimension results in divergent quantum corrections which are
localised at the orbifold points. We are thus forced to introduce
brane Lagrangians at these points even if we did not start out
with them, with couplings of the same form as the divergences (so
that renormalizing the couplings absorbs the divergences). By the
symmetry at the branes, we need four times as many couplings
$1/g^2_i$ as there are semi-simple factors in $H$, to multiply
separately the generated terms $(\F^\ha_{\mu\nu})^2$ and
$(\F^\ha_{\mu5})^2$ on each boundary (see \sec{GHU} for an
example). Using the Lagrangian in the form \eq{bulk}, and for
example working in background field gauge \cite{Abbott} \cf
\fig{fig:selfE}, it is easy to see that the couplings then run
according to $\mu\partial_\mu(1/g^2_i)=b_i/(8\pi^2)$ where these
$b_i$ are combinations of group theory factors, yielding numbers
say, in the range $O(1)$ to $O(10)$, but whose value depends on
the details (in particular the orbifolding and the fermions we are
about to introduce).

\FIGURE{
\epsfig{file=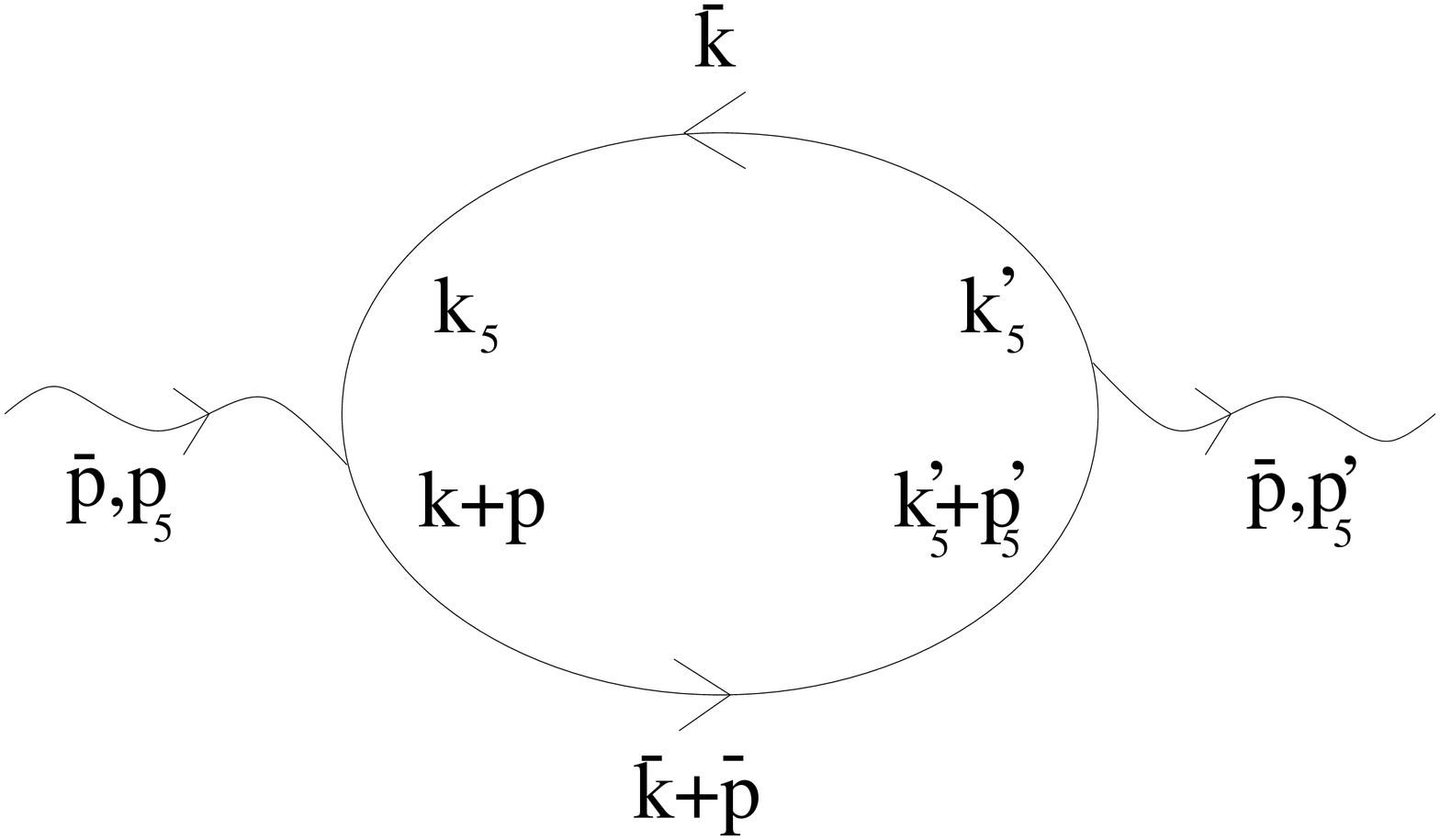,scale=.25}
\caption{One-loop contribution to the $\beta$ function in
background field gauge. The internal line stands for all
components, including ghosts.} \label{fig:selfE} }

Note that in computing these effects we work in 5 dimensions: the
localised divergences only arise from integrating over the
four-dimensional loop momentum $\bk^\mu$, whilst the 5th component
only participates in the Umklapp process $\sim \delta_{2k_5,p_5\pm
p'_5}$ which in position space leads to the brane localised delta
functions $\delta(y)+\delta(y-\pi R)$ \cite{Georgi}.\footnote{The
loss/gain of KK momentum together with possible reflection of the
momenta is analogous to Umklapp interactions with lattices in
condensed matter. Indeed in reality the branes absorb the change
in momentum.} We would get the wrong answer if we worked in
$4+\epsilon$ dimensions, thus expanding the $\bk$ momentum
integral around three dimensions. Indeed in this case the diagrams
would all be ultraviolet finite, and all infrared finite, except
for the $k_5=0$ term which is linearly IR divergent for vanishing
$p$. Whether we work in 5 dimensions or expand around 4 dimensions
has to be determined by the effect being calculated: we need to
choose to expand about 4, if and only if it is the critical
dimension for the effect in question, \ie the dimension at which
logarithmic divergences appear. At one loop, this clearly decides
the issue. Perhaps at two loops and higher, there can be an effect
that can only be properly understood by considering its
divergences both in 5 and around 4 dimensions? We leave this
question for the future.

Note that it is not possible to argue that the $1/g^2_i$ are much
less than $b_i/(8\pi^2)$, because the result would then be
strongly $\mu$ dependent. Indeed in this case, changing the value
of $\mu$ by a factor of two, would return the couplings to
$1/g^2_i(\mu)\sim b_i/(8\pi^2)$. We thus conclude that the brane
couplings are $O(10^{-1})$ or greater.

Even if we were to limit the appearance of these brane kinetic
terms by arranging the $b_i$ to be very small or cancel (at one
loop or to any number of loops), it is not possible to set
$1/g^2_i$ to zero if there are other non-vanishing brane
interactions involving $\A$, without generating further
divergences, as we explain below.

Since these couplings $1/g^2_i$ do not respect the full gauge
symmetry $G$, Grand Unification in a single gauge group is not
really possible in these scenarios (even if some of its features
such as charge quantization can be preserved \cite{Larry}). For
this reason, and the fact that lack of renormalizability is less
of a problem in any case at Grand Unified scales (\cf \sec{INT}),
we do not pursue the possibility of renormalizable extra
dimensional Grand Unified Theories further here.

\FIGURE{
\epsfig{file=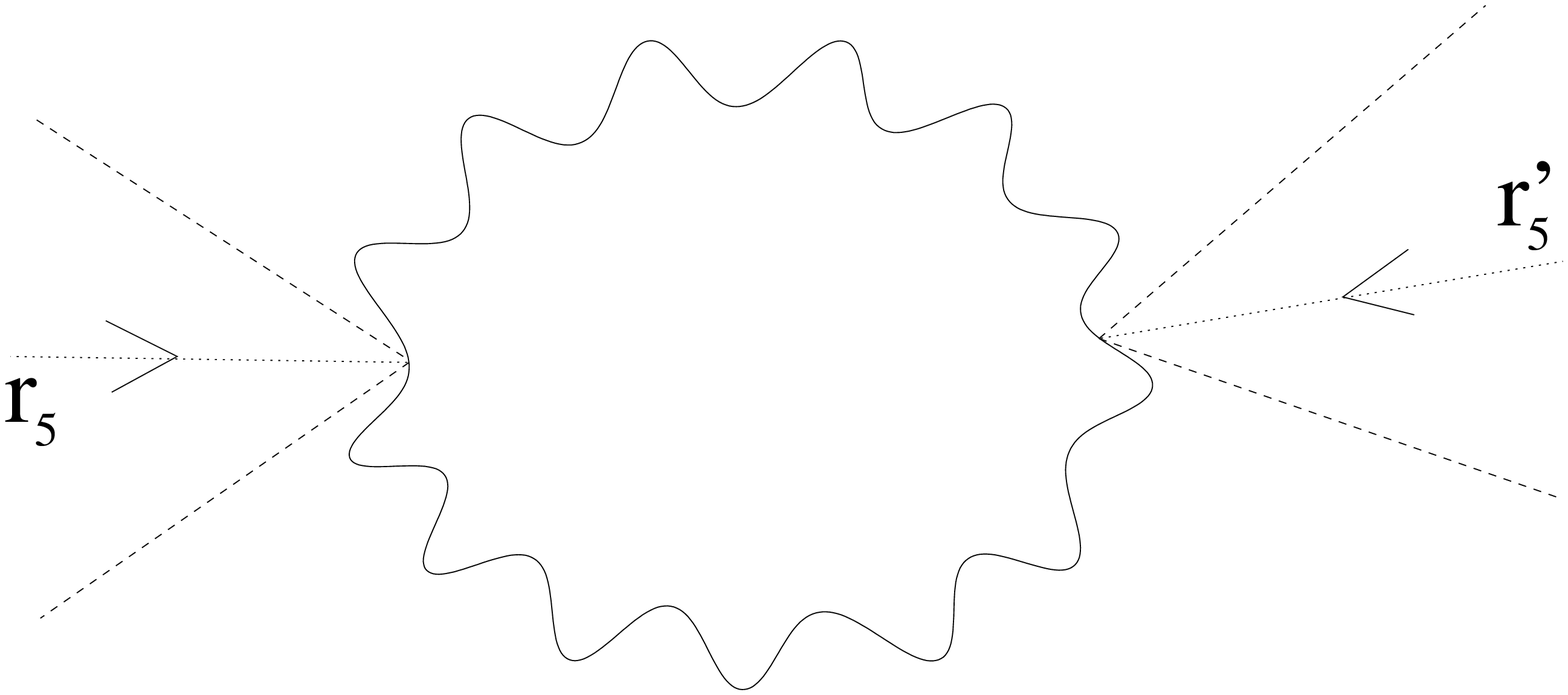,scale=.25}
\caption{Contribution to brane-localised $\phi^4$ interaction. The
dashed lines are external $\phi$s. The dotted lines are insertions
of external $y$-momentum from \eq{branefn}.} \label{fig:seagull} }

To see that it is not possible to switch off brane kinetic terms
for $\A$ in the presence of other non-vanishing brane interactions
for $\A$, consider adding an $H$ invariant scalar field $\phi(x)$
to the brane(s). This will generate from the kinetic term in
particular a brane interaction of the standard ``seagull'' type
\be
\label{seagull}
\delta(y)\A\A\phi\phi
\ee
to be added to the overall Lagrangian density \eq{bulk}. The
effect we are about to describe happens for any field propagating
in the bulk and any interaction, so for the moment we will ignore
the Lorentz and colour indices, treating $\A$ as though it were a
scalar field. Since this is a small-distance effect, it occurs
just as well in infinite flat bulk dimensions with a single brane
at $y=0$.

Now the one-loop diagram indicated in \fig{fig:seagull} generates
a local divergent correction to the brane $\phi^4(x)$ interaction.
The delta function in \eq{seagull} is easily taken into account if
we regard it as a background field \be \label{branefn} \phi_b(x,y)
= \delta(y) \ee in a five-point interaction, as indicated in
\fig{fig:seagull}. We see that the local interaction generated is
actually \be \label{dsquared} \phi_b^2\phi^4 \ee which is not well
defined with the identification \eq{branefn}.

Part of the problem of course lies in our assumption that the
brane is infinitely thin and infinitely heavy. In a fully
realistic situation, the brane would have a finite mass which we
identify with the scale of quantum gravity $\Lambda$, and a form
factor
\be
\label{formfactor}
\phi_b(x,y) = f_b(y),
\ee
where $f_b$ becomes a delta function only in the limit
$\Lambda\to\infty$. If $\A$ were not a gauge field, we could
indeed replace $\delta(y)$ by \eq{formfactor}. Since for functions
smooth at distances $1/\Lambda$,
\be
\label{ident}
\phi_b^2 \equiv c \Lambda\phi_b,
\ee
(where $c$ is some constant of order 1) this extra divergence
could be absorbed into the couplings. However, in our gauge
theory, the replacement of $\delta(y)$ by a form factor would be
fatal: in order for \eq{seagull} to remain gauge invariant, $\phi$
has to be extended so that it depends on the bulk dimensions. This
would mean we would not have the option to have brane-localised
fields, severely limiting not only the model-building
possibilities but also the possibility to keep the bulk theory
asymptotically free in four dimensions and thus supportable by an
ultraviolet fixed point in $4+\epsilon$ dimensions. An alternative
way to proceed might be to use \eq{branefn} and promote relations
such as \eq{ident} to identities. At this stage it is not clear if
the model can then be made fully renormalizable.

Fortunately we do not have to pursue these directions further,
since the divergence does not appear if we keep a brane kinetic
term for $\A$. To see this, again for simplicity we treat $\A$ as
a single component real scalar field. After a rescaling the
kinetic terms yield the inverse propagator \be \label{invprop}
\Delta^{-1} = -\,\partial^M\partial_M-{2\over m}
\,\delta(y)\,\partial^\mu\partial_\mu, \ee where $m\equiv
2g_i^2/g^2$ is the ratio of couplings and has dimensions of mass.
The corresponding propagator has already been derived in the
literature (see \eg \cite{0005016}). However, to discuss
divergences it is helpful to write it completely in momentum
space. This is easily found directly as follows. In Euclidean
momentum space, \be \Delta^{-1}(\bp,p_5;\bq,q_5) = p^2\delta(p-q)
+ 2\,{\bp^2\over m} \,\delta(\bp-\bq), \ee where $\delta(k)$
($\delta(\bk)$) means the $D=5 (4)$ dimensional delta function
multiplied by $(2\pi)^{D}$. Substituting the ansatz, \be
\Delta(\bp,p_5;\bq,q_5) =
a(p)\delta(p-q)+b(\bp,p_5,q_5)\delta(\bp-\bq), \ee where $a(p)$
and $b(\bp,p_5,q_5)=b(\bp,q_5,p_5)$ are functions to be determined
one finds \be \Delta(\bp,p_5;\bq,q_5) = {\delta(p-q)\over p^2} -
\,{2\bp^2\delta(\bp-\bq)\over p^2(\bp^2+q_5^2)(m+|\bp|)}\,. \ee
(The modulus term $|\bp| = \sqrt{\bp^2}$ arises from integration
over $q_5$.)

Taking account of momentum conservation in the usual way, one
finds the contribution \fig{fig:seagull} is proportional to
\be
\int d^4x\, d^4x'\, \phi^2(x,0)\,\phi^2(x',0) \int\!\!
{d^4\bp\over(2\pi)^4}\,\, {\rm e}^{i\bp.(x-x')}\int\!
{d^4\bq\over(2\pi)^4}\, \Delta(\bq) \Delta(\bq+\bp)\,,
\ee
where the effective four-dimensional propagator
\be
\label{effprop}
\Delta(\bp) = {1\over\bp^2+m|\bp|}
\ee
comes from integrating over the injection of $r_5$ ($r'_5$) with
coefficient 1, that arises at each vertex from \eq{branefn}.

Note that the effective propagator results in no worse ultraviolet
divergences than a four dimensional theory completely confined to
the brane. It is clear that all one-loop diagrams of this sort
will be regulated in the same way and thus no terms of the form
\eq{dsquared} are generated. Intuitively one can understand what
is happening as follows. For very small distances on the brane,
the $\delta(y)$ term in \eq{invprop} dominates. In this regime
everything is confined to the brane and we have just a four
dimensional theory. The five dimensional coupling constant $g$,
through $m$, provides a cross-over scale in the effective
propagator \eq{effprop}, so that only for distances larger than
$1/m$ do we feel the effects of the full 5 dimensions. In fact
what we have is the gauge theory analogue of the ``DGP'' model
\cite{0005016}, where these effects are discussed in a
gravitational context, a neat derivation of precisely \eq{effprop}
appearing in ref. \cite{0303116}. This intuition gives us
confidence that this ultraviolet regularisation effect works to
all loop orders.

Finally, note that the above considerations imply that all the
$\beta$ functions of the brane-localised fields (here that of a
brane scalar field) must be singular in the limit $1/g^2_i\to0$,
in which the brane kinetic terms are turned off.

Note that the gauge field components $\A^{a'}_5$ are even under
$P$. Their lowest KK modes thus behave like scalars as far as the
low energy four-dimensional theory is concerned. There is nothing
in the four-dimensional theory that prevents them from gaining a
mass. From the five dimensional point of view this is not allowed
by gauge invariance. However, an effective action can and will
appear for the non-local operator \be \label{Polyakov} \Phi_P(x)=
\Phi(x,2\pi R), \ee the Wilson line that winds around the compact
dimension, where \be \Phi(x,y) = {\cal P} \exp\, -i \int_0^y\!\!\!
d{\tilde y}\, \A_5(x,{\tilde y}) \ee is the Wilson line from
${\tilde y}=0$ to ${\tilde y}=y$, located at $x$ (and ${\cal P}$
stands for path ordering). Since under gauge transformations, \be
\label{Phigauged} \delta\Phi(x,y) =
i\,\Omega(x,0)\,\Phi(x,y)-i\,\Phi(x,y)\,\Omega(x,y), \ee
$\Phi_P(x)$ transforms homogeneously as an adjoint scalar at
$(x,0)$.

Since $\Phi_P$ contains a power series in $\A^{a'}_5$, the gauge
invariant effective potential $\tr V_{eff}(\Phi_P)$ not only
generically results in masses for the lowest KK mode of the
components $\A^{a'}_5$, but can also result in spontaneous
symmetry breaking: $<\!\A^{a'}_5\!>\,\ne0$. This is the Hosotani
mechanism \cite{Hosotani}. In the spontaneously broken phase,
although we can by a gauge transformation set $\A_5=0$, this is at
the expense of a non-trivial Scherk-Schwarz twist in periodicity
conditions for $\A$ \cite{13} (as we review in \sec{GHU}) which
also breaks the gauge group. Meanwhile, $V_{eff}$ and the low
energy physics as described through it, are completely unaffected
by this gauge transformation. Note that the mass and more
generally the effective potential is protected from divergent
corrections: the effect is non-local in the full five-dimensional
space and thus quantum corrections are cutoff by the
compactification scale $1/(2\pi R)$.

Clearly an extremely attractive possibility now arises, namely to
regard the Higgs scalar field as the components $\A^{a'}_5$
\cite{Hosotani,Kubo:2001zc,Haba:2002py,MasieroS3,AntoniadisBQ,HallNS,N1,BurdmanNomuraSUSY,GIQResidualGaugeSym,GIQ6D,S3},
for if we can bring such a model into accord with phenomenological
constraints, whilst keeping it renormalizable, we will have solved
the hierarchy problem. This is the possibility we will consider in
more detail in \sec{GHU}. We will use the framework developed
there to make general comments on model building constraints that
arise when we consider couplings to fermions.

\section{The simplest gauge-Higgs unification model}
\label{GHU}

The simplest possibility for the unified gauge group is $G=SU(3)$
\cite{SU3}.\footnote{This is really $SU(3)\times SU(3)$. The
second $SU(3)$ is for strong interactions and is readily
incorporated in the bulk. Since the real problems lie with the
weak interactions we will not discuss colour further.} With the
generators $T^a=\lambda^a/2$, where the $\lambda^a$ are the
Gell-Mann matrices, we can decompose $\A_M$ as
\be
\label{decomposeA}
\A_M = \pmatrix{W_M & H_M/\sqrt{2}\cr H^\dagger_M/\sqrt{2} &0}+B_M
T^8.
\ee
Here, $(W_M)^i_j$, with $i,j=1,2$, contracted into $T^\ha$
($\ha=1,2,3$), are the gauge bosons associated to the top left
$SU(2)$ subgroup. This will be identified with the $SU(2)_W$ part
of the Standard Model gauge group. Similarly $B_\mu$ will be
identified with the $U(1)_Y$ gauge boson, while $H^i \equiv H^i_5
= \sqrt{2} (\A_5)^i_3$ is to be identified with the $SU(2)_W$
doublet Higgs. We implement the restriction of $SU(3)$ to
$SU(2)_W\times U(1)_Y$, by identifying the parity operator $P$ as
mapping the third component of the $SU(3)$ fundamental
representation to minus itself, whilst leaving components $i=1,2$
alone. Under this map indeed, $W_\mu$, $B_\mu$ and $H$ are even
and thus at this stage will have massless KK modes, whilst $W_5$,
$B_5$ and $H_\mu$ are odd and thus only have KK masses of order
the compactification scale or greater.

The radiatively generated potential for $\Phi_P$, will then
spontaneously break the model, effectively giving $H$ a vacuum
expectation value, via the Hosotani mechanism as outlined in the
previous section. Using the remaining global symmetry we can
ensure the vacuum expectation value takes the form $<\!H\!>=
(0,v)/\sqrt{2}$ for some real $v$. Thus $SU(2)_W\times U(1)_Y$
will be broken down to electromagnetism, just as happens in the
Standard Model.

Since
\be
\label{t8h}
[T^8,H] = {\sqrt{3}\over2} H,
\ee
we already have a problem however. If we were to put the coupling
$g$ back in its normal place in the covariant derivative, we would
want to identify $g T^8$ with $g' Y/2$ where $Y$ is the
hypercharge and $g'$ the associated coupling. Comparing to the
Standard Model Higgs for which $Y=1$, we get from \eq{t8h}, that
$g'=\sqrt{3}g$ and thus we `predict'
$\sin^2\theta_W=g'^2/(g^2+g'^2)=3/4$, a phenomenological disaster.

However, we now recall that we necessarily have the following
couplings on the branes: \be \Delta\LL = \delta(y)\, \LL_1 +
\delta(y-\pi R)\,\LL_2, \ee where, writing out the non-vanishing
parts of $\F_{\mu5}^2$ and similarly the semi-simple components of
$\F_{\mu\nu}^2$, \be \label{brane} \LL_\alpha =
-{1\over2w^2_\alpha}\,\tr\,W_{\mu\nu}^2
-{1\over4b^2_\alpha}\,B^2_{\mu\nu} +{1\over
h^2_\alpha}\,|\nabla_\mu H-H'_\mu|^2. \ee Here we have defined
$H'_\mu = \partial_5 H_\mu$, the brane couplings $w_\alpha$,
$b_\alpha$ and $h_\alpha$, and the field strengths $B_{\mu\nu} =
\partial_\mu B_\nu - \partial_\nu B_\mu$ and $W_{\mu\nu} = i[D_\mu,D_\nu]$,
where $D_\mu = \partial_\mu - i W_\mu$. Note that \be \label{covH}
\nabla_\mu H = \partial_\mu H - iW_\mu H -i {\sqrt{3}\over2}B_\mu
H, \ee as a consequence of \eq{t8h} and the restriction to the
branes.

Recall that the couplings $1/w^2$, $1/b^2$ and $1/h^2$ are
$O(1/10)$ or larger. (The symmetric point where they have the same
value on each brane will typically be broken when fermions are
included.) In the presence of such couplings \eq{brane},
spontaneous symmetry breaking deforms the lowest mass modes for
the weak vector bosons $W^\pm_\mu$ and $Z_\mu$ so that they are no
longer simply constants in the $y$ direction (in any gauge)
\cite{S3}.

However, this effect is controlled by the ratio of scales
$\theta/2\pi = Rv/2$ \cite{S3}, vanishing as $\theta\to0$. This
ratio appears in
\be
\label{Wiloop}
<\!\Phi_P\!\!>\ =
\pmatrix{1&0&0\cr0&\cos\theta&-i\sin\theta\cr
0&-i\sin\theta&\cos\theta}.
\ee
The physics is invariant under
$\theta\mapsto\theta+2\pi$. Defining $\theta$ to be in the
fundamental domain $[0,2\pi)$, the natural theoretical expectation
is that $\theta/2\pi\sim 0.5$ if spontaneous symmetry breaking
takes place. However, indirect limits require the compactification
scale $1/R = a\TeV$ where $a\gsim 2 - 5$ or greater \cite{18}.
Identifying $v$ for the moment with the Standard Model Higgs'
vacuum expectation value, we require $\theta/2\pi\lsim0.06$. In
this case we can ignore the deformations in the first
approximation and take the lowest mass modes for $W^\pm_\mu$,
$Z_\mu$ and $H$ to be constant in the $y$ direction.

The form of the effective Lagrangian for these modes is then
\bea
&&2\pi
R\left[-{1\over2g^2}\tr\,W_{\mu\nu}^2-{1\over4g^2}B_{\mu\nu}^2
+{1\over g^2}|\nabla_\mu H|^2\right]\cr
&&-{1\over2}\tr\,W_{\mu\nu}^2\sum_\alpha{1\over w^2_\alpha}
-{1\over4}B_{\mu\nu}^2\sum_\alpha{1\over b^2_\alpha} + |\nabla_\mu
H|^2\sum_\alpha{1\over h^2_\alpha}\,,
\eea
where \eq{covH} holds but now with $y$-constant modes. Thus
clearly if we define
\bea
\label{r1}
{1\over g^2_2} &=& {1\over2\pi^2\ta} +{1\over w_1^2} + {1\over
w_2^2}\\
\label{r2}
{3\over g^2_1} &=& {1\over2\pi^2\ta} +{1\over b_1^2} + {1\over
b_2^2}
\eea
where we have used the Wilsonian dimensionless bulk coupling
\eq{ta} evaluated at $\mu=1/R$ (and $D=5$, $N=3$), \viz $\ta=
g^2/4\pi^3 R$, and redefine
\be
\label{Smredef}
W_\mu = g_2\, \tW_\mu,\qquad B_\mu=
{g_1\over\sqrt{3}}\,\tB_\mu,\qquad H= \tH \left( {1\over2\pi^2\ta}
+{1\over h_1^2} + {1\over h_2^2} \right)^{-1/2}\ ,
\ee
to put couplings back in their usual places, we get back precisely
the relevant part of the Standard Model Lagrangian
\be
\label{SM}
-{1\over2}\tr\,\tW_{\mu\nu}^2-{1\over4}\tB_{\mu\nu}^2+|\nabla_\mu
\tH|^2
\ee
where in particular $\tW_{\mu\nu}$ is now defined as expected in
terms of $D_\mu = \partial_\mu - ig_2\tW_\mu$, and the full
covariant derivative
\be
\label{rcov}
\nabla_\mu  = \partial_\mu  - ig_2\tW_\mu  -i g_1{Y\over2}\tB_\mu
\ee
defines a hypercharge
\be
\label{Y}
Y= {\rm diag}(1,1,-2)/3
\ee
which gives the Higgs $Y=1$ as expected.

Taking into account the rescaling to $\tH$, we see that $v$ is
actually defined in terms of the Standard Model Higgs' expectation
value $\tv$ by
\be
\label{r3}
R\ \tv = {\theta\over\pi}\sqrt{ {1\over2\pi^2\ta} +{1\over h_1^2}
+ {1\over h_2^2} }.
\ee

These equations allow us to make several straightforward but
important conclusions. Firstly, in the limit that $\theta$ is
sufficiently small to neglect deformations of the lowest KK
wavefunctions, we regain from \eq{SM}, the custodial symmetry of
the Standard Model and thus determine the $\rho$ parameter to be
one at tree level, as required. Secondly, using the experimentally
determined numbers \cite{Hagiwara}, the bound $1/R = a$ TeV and
the fact that $1/w^2$, $1/b^2$ and $1/h^2$ are $\gsim1/10$ we can
place further bounds on the values of the parameters.

Thus from \eq{r1}, $1/2\pi^2\ta<2.3$, and with natural values for
$w_\alpha$, we have $\ta\gsim0.02$. This is clearly perturbative
and an order of magnitude smaller than the typical fixed point
value (\cf \sec{RED}). At energies higher than $1/R$, $\ta$ will
run according to \eq{beta}, reaching the non-perturbative physics
associated with the fixed point at energy scales $\sim 10/R$.

On the other hand we clearly have an upper bound on
$1/w^2_1+1/w^2_2$ of $2.3$. From \eq{r2} the strict upper bound on
$1/2\pi^2\ta$ implies a {\sl lower} bound on $1/b^2_1+1/b^2_2>21$.
We therefore require very large $U(1)_Y$ kinetic terms on the
boundary. Perhaps these arise naturally from non-perturbative
physics close to the ultraviolet fixed point $\ta_*$, bearing in
mind that Abelian gauge fields would be separately
non-renormalizable in the bulk. Of course phenomenologically, the
origin of the large values is the discrepancy between the `bulk'
$\sin^2\theta_W=3/4$ and the experimental one.

The Lagrangian given by \eq{bulk} and \eq{brane} is very similar
to the bosonic sector of the Lagrangian considered in ref.
\cite{S3}, except that Scrucca \etal introduce an extra bulk
$U(1)$ gauge field to allow $\sin^2\theta_W$ to be set to its
experimental value (and do not introduce the $h_\alpha$ terms). We
do not have the option of including a bulk $U(1)$ gauge field
since it is not renormalizable within the $\epsilon$ expansion.

Finally, \eq{r3} implies that $\theta/2\pi>0.08/a$ with natural
values for $1/h^2_\alpha$. Therefore phenomenologically preferred
values of $\theta$ are consistent with our approximation.

To allow for larger $\theta$, we have to increase $\ta$. Thus
$\ta\sim0.1$ (implying $1/w^2\sim1$) allows for
$\theta/2\pi\sim0.18/a$, but to get the theoretically natural
values of $\theta/2\pi\sim0.3$ \cite{S3} requires $\ta\sim1.2$.
Such a large value would probably (depending on matter content)
imply that the fixed point $\ta$ is being approached from the
right (\ie from the region $\ta>\ta_*$). There is no problem of
principle with this happening but in particular, large higher
order corrections in the $\epsilon$ expansion mean that we could
then no longer trust these simple formulae. However, in any case
we are back in a regime where significant distortions from
Standard Model relations will be found. We should note that there
are not enough parameters in the model in this regime to tune away
the resulting anomalous triple gauge couplings \cite{S3,tgbs},
tune $\rho$, $\sin^2\theta_W$ and $m_Z$ and the effective brane
couplings to matter fields \cite{S3} to their correct values. Even
if we could manage this, it would be an unsatisfactory accident
since custodial symmetry has been badly broken.

In summary, for the bosonic sector of the model to be
phenomenologically acceptable we must have large values of
$1/b^2\sim10$, and a much smaller $\theta$ than we would find
without some special mechanism. Indeed we can choose a small
$\ta>0.02$ and natural values of $1/w^2 \sim 1/h^2 \sim 1/10$, in
which case we have to arrange the model to dynamically determine
$\theta/2\pi\sim0.04-0.09$.

Before turning to the introduction of fermions we discuss briefly
the peculiar remnants of gauge invariance operating on the brane.
Decomposing a gauge transformation similarly to \eq{decomposeA},
as
\be
\Omega = \pmatrix{\omega & \phi\cr\phi^\dagger &0} + \beta T^8,
\ee
we have that $\omega$ and $\beta$ are $P$ even, while $\phi$ is
$P$ odd. Thus $\omega$, $\beta$ and $\phi'=\partial_5\phi$ survive
on the branes whilst $\partial_5\omega$, $\partial_5\beta$ and
$\phi$ vanish on each brane. It follows that the brane Lagrangians
\eq{brane} are invariant under the surviving, or ``remnant''
\cite{GIQResidualGaugeSym}, gauge symmetries:
\bea
\label{remnants}
\delta W_\mu &=& [D_\mu,\omega] \cr
\delta B_\mu &=& \partial_\mu\beta \cr
\delta H &=& \phi' + i\omega H + i{\sqrt{3}\over2}\beta H \cr
\delta H'_\mu &=& i\omega H'_\mu + i{\sqrt{3}\over2}\beta H'_\mu +
\nabla_\mu\phi',
\eea
where $\nabla_\mu\phi'$ is \eq{covH} with $H$ replaced by $\phi'$.

As noted in ref. \cite{GIQResidualGaugeSym}, the shift symmetry
$\phi'$ protects against brane mass terms appearing for the Higgs.
It is interesting also to note that if it were not for the bulk
Lagrangian \eq{bulk}, the brane Higgs kinetic term in \eq{brane}
would be trivial since it could be gauged away by a finite
$\phi'=H$ transformation in \eq{remnants}, $H'_\mu$ playing the
r\^ole of an auxiliary field. However, this term becomes
non-trivial when considered as part of the full Lagrangian:
although we can still gauge the brane Higgs kinetic terms away,
this is a gauge choice which in general conflicts with the need to
make other gauge choices (for example the background Feynman gauge
used typically to compute the radiative potential). Furthermore,
$H'_\mu$ is of course no longer an auxiliary field but part of the
bulk degrees of freedom evaluated at the orbifold points.

In order to complete a realistic description we need to introduce
quarks, and leptons. For example consider initially, the top and
bottom quarks. We write their $SU(2)_w\times U(1)_Y$ ${\bf
2}_{1/6}$, ${\bf 1}_{2/3}$, ${\bf 1}_{-1/3}$ representations as
$Q_L = (t_L,b_L)$, $t_R$ and $b_R$. As usual in gauge-Higgs
unification models, we cannot introduce these as bulk fermions
because the interactions would be flavour-symmetric and have the
wrong hypercharges, following from \eq{Y}. If we introduce them as
fields that live only in the brane(s), then we can simply assign
them the correct hypercharges as according to \eq{rcov}, since
only the restricted symmetry \eq{remnants} is active there.
However, \eq{remnants} as well as including the standard
$SU(2)_W\times U(1)_Y$ gauge transformations (after use of
\eq{Smredef} and defining ${\tilde\beta} = \sqrt{3}\beta$)
includes the shift symmetry generated by $\phi'$ which forbids the
Yukawa interactions, for example under the shift symmetry we have
$\delta({\bar Q}_LHb_R) = {\bar Q}_L\phi'b_R$.

(We can cancel this by postulating heavy mirror fields, \eg a
${\bf 2}_{1/6}$ $Q_R$, adding appropriate mass terms $M{\bar
Q}_LQ_R$ (plus c.c.), and defining $\delta Q_R \sim\phi'b_R$, and
so on, resulting in a see-saw mechanism with the lightest states
to be identified with the quarks. However we then need kinetic
terms for the $Q_R$. The terms generated by the shift
transformations acting on this can be cancelled by introducing an
appropriate interaction with $H'_\mu$ and appropriate
transformations into heavy partners for $b_R$ and $t_R$.
Unfortunately then more fields are needed to in order to cancel
new violations of the shift symmetry and so on. It appears that it
is not possible to find a closure of the symmetry which is
non-trivial, linear and finite dimensional. However, since we are
already introducing an infinite number of fields through the KK
excitations, it could be worthwhile to pursue the possibility of
infinite dimensional representations of the shift symmetry. Note
that if these could be constructed, the shift symmetry would still
protect the Higgs from gaining a brane potential, but allow
standard Yukawa interactions for the real quarks and leptons.)

This problem has been circumvented in the literature by taking
instead the Wilson line $\Phi$, using $\Phi_P$ for coupling
doublets and singlets on the brane at $y=0$ (analogously a Wilson
line that wraps once round the compact dimension but starting at
$y=\pi R$, for coupling both representations at $y=\pi R$), or
$\Phi(\pi R)$ for coupling fermions on one brane to fermions on
the other \cite{N1}. This can be done because from \eq{Phigauged},
\be
\label{homohiggs}
\delta\,\Phi^i_{\ 3}(x,y)= i \omega(x,0)\Phi^i_{\ 3}(x,y) +
\left[{i\over6}\tb(x,0)+{i\over3}\tb(x,y)\right]\Phi^i_{\ 3}(x,y)
\quad {\rm for} \quad y=0,\pi R.
\ee
Therefore $\Phi^i_{\ 3}(x,y)$ transforms homogeneously like the
Higgs in \eq{remnants} but without the shift symmetry. From the
above we see we can actually provide Yukawa couplings between
different branes as ${\bar Q}_{Li}(x,0)\Phi^i_{\ 3}(x,\pi
R)b_R(x,\pi R)$, but $U(1)_Y$ invariance requires the standard
coupling for the charge conjugates $Q_R^c$ and $t^c_L$ to be on
the same brane. We may similarly provide couplings for the other
quark families, and also for the leptons, where we just use the
fact that for $\Phi_P$,  \eq{homohiggs} with $y=0$ is the usual
Higgs transformation.

Such Wilson line interactions can arise from integrating out heavy
bulk fields \cite{N1,S3,HallNS} and also arise in String
compactifications \cite{ibanez,N1}. There is no problem here and
later with brane localised anomalies since they may be cancelled
by an appropriate bulk Chern-Simons action \cite{anomaly}. However
as noted by ref. \cite{N1} such Yukawa interactions produce new
brane-localised divergences. In particular they will give
quadratically divergent contributions to the Higgs mass
\be
\label{yuk}
\sim \Lambda^2\Phi^i_{\ 3}\Phi^{\dagger3}_{\ \ i}
\ee
just as in the Standard Model. Of course this destroys the purpose
of the model as a solution to the hierarchy problem.

The reason is immediately clear because the $y$ degree of freedom
plays no r\^ole here for the brane localised fermions. As far as
they are concerned $\Phi^i_{\ 3}$ looks just like the Standard
Model Higgs. Indeed we would also have to add brane kinetic terms
and a brane-localised quartic potential for $\Phi^i_{\ 3}$ to
absorb logarithmic divergences and make the model
renormalizable.\footnote{It is not clear to us whether the
resulting framework is renormalizable in 5 dimensions, however if
it is not, then the right approach is to consider these effects
expanded around 4 dimensions, where it is renormalizable - see our
comments below \eq{effprop}.}

In ref. \cite{N1}, the authors are interested only in the little
hierarchy problem, and circumvent the further difficulty \eq{yuk}
for the top by adding a new colour triplet fermion  (for us this
would be a ${\bf 1}_{-1/3}$) so as to complete the representation
$Q_L$ to a representation of $SU(3)_W$. Then the summation over
$i$ in \eq{yuk} in fact runs over the complete representation
($i=1,2,3$) becoming just a contribution to the vacuum energy
($\Phi\Phi^\dagger=1$). The divergences from the other fermions
are small enough to ignore in this scenario since the model anyway
has a low effective cutoff $\Lambda$ and the corresponding Yukawa
couplings are much smaller. With a Planck mass cutoff, we do not
have this option, even for neutrinos and their very low masses,
and clearly it is phenomenologically unacceptable to add
hypercharged but $SU(2)_W$ singlets to fill out representations of
$SU(3)_W$ for the other quark families and the leptons.

The appearance of divergences from such non-local operators seems
to go against the standard lore that divergences arise from local
interactions only. There is no contradiction however. These Wilson
line interactions are local in space-time ($x$ space). This is an
approximation which is valid only in the limit of infinite string
tension (in the case of String Theory compactifications) or
infinite mass (in the case of massive bulk fields). In truth one
should integrate over different locations $x$ and $x'$ where the
ends of the Wilson line meet the fermions, with the displacement
$x-x'$ being weighted by a form factor of width $1/M$, where $M$
is the heavy scale. These Wilson line interactions are then indeed
ultraviolet finite, quantum corrections being naturally cutoff at
$M$.

Since our model is supposed to be valid up to the Planck mass
$\Lambda$ however, we are forced to describe the dynamics that
results in such smeared Wilson line interactions (unless
$M\gsim\Lambda$ in which case we are back at square one). We are
left with the remaining concrete alternative in the literature
which is indeed to generate the Yukawa interactions through
couplings of bulk and boundary fields \cite{S3,BurdmanNomuraSUSY}.

We follow closely the ideas of ref. \cite{S3}. Thus we introduce a
pair of bulk fermions $\Psi_1(x,y)$ and $\Psi_2(x,y)$ (necessarily
Dirac fermions since they are in 5 dimensions) with opposite
$y$-parity so that a parity invariant mass term
$M{\bar\Psi_1}\Psi_2$ (plus c.c.) can be built. Using \eq{Y}, we
choose a set of representations of $SU(3)$ so that the Standard
Model fermions, which exist only on the branes, can all couple to
them and thus all the Standard Model fermions (including the
neutrinos) can get masses from the effective Wilson line
interactions that will be generated. Thus we can use a
$\Psi_\alpha$ pair in the fundamental (${\bf3}$) representation so
that $\Psi^i$ couples to $Q_L$ while $\Psi^3$ couples to $b_R$.
This is the minimal representation that will do the job. The
minimal representation that yields components with the right
hypercharge to couple to $t_R$ is the {\bf 6}, where we use the
$33$ component to couple to $t^c_L$. (The {\bf 6} will also couple
to $Q^c_R$.) We deal with the other quark families in precisely
the same way. Charged conjugated lepton doublets can couple first
to the adjoint ({\bf 8}) representation; the charge conjugated
right handed electron, muon and tau couple to the $333$ component
of a {\bf 10}, and finally the right handed neutrinos can couple
to the ${\bf 1}_0$ in the {\bf 8} representation $\Psi_\alpha$s
already introduced.

As we have already discussed, something special is needed to get a
low value $\theta/2\pi \sim0.04-0.09$. In ref. \cite{0401183} it
was shown that this is possible by adding of order 10 bulk fields
in a mixture of fundamental and adjoint representations. They used
scalars as well as fermions. Although scalars are problematic for
us, they do not appear to be especially required. These authors
assume that the correct value of $\sin^2\theta_W$ can be realised
by wall localised kinetic terms. As we have shown, this can indeed
be achieved. Somewhat similarly, the authors of ref. \cite{S3}
manage to reduce the minimum to $\theta/2\pi\sim0.096$ by adding
bulk fermions in large (rank 8) symmetric representations. They
note however that this will lead to electroweak corrections
enhanced by large group theoretical factors resulting in the scale
at which the bulk weak coupling becomes non-perturbative being
lowered (their cutoff by application of NDA). Here we do not have
this problem.

For quite separate reasons we have also introduced a large number
of bulk fermions. It would be very interesting to see if the
menagerie of representations we have had to introduce to induce
Yukawa matrices for the Standard Model fermions, also turned out
to give values of $\theta$ in the right range. However, we now hit
a severe problem if we want to preserve the renormalizability of
the model. The contributions of the bulk fermions to the bulk
$\beta$ function is\footnote{The factor 8 comes from the usual
$4/3$, the two bulk fermions and the three families. Note that
boundary fields and couplings make no contribution to the bulk
$\beta$ function.}
\be
3\beta_0 = - 8 \sum_R T_R = -4\, (1+5+6+15) = -108
\ee
completely overwhelming the contribution of $+11$ from $\A_M$.
Thus the bulk theory is no longer asymptotically free in four
dimensions and cannot be supported by an ultraviolet fixed point
in $4+\epsilon$ dimensions.

This constraint would appear to rule out renormalizable models
based on a bulk $SU(3)$ Yang-Mills theory. The obvious route to
try to make further progress would be to consider larger gauge
groups, thus increasing the gauge-field contribution to $\beta_0$,
while also allowing more Standard Model fermion representations to
couple to the same bulk fermion representations. (For example
above, the {\bf3} is shared by $Q_L$ and $b_R$, while the right
handed neutrinos couple to the same representation as the lepton
doublets.) This results in considering a kind of grand unification
of a quite different sort from the standard four dimensional
cases, but with some similar properties, for example charge
quantization and mass relations. However, we also clearly need to
consider more complex orbifolds, and/or deal with multiple Higgs
vacuum expectation values arising from the Hosotani mechanism (see
ref. \cite{0401185} for such a study).

\section{Summary and conclusions}
\label{CON}

Extra dimensional field theories have the potential to solve many
of the enduring mysteries of theoretical particle physics. However
approaches that directly address weak-scale physics and in
particular the hierarchy problem, suffer from a severe drawback in
that they are not renormalizable, at least as conventionally
envisaged. This results in an irreducible uncertainty of typically
$\sim$ 1\% in any predictions following from these models, and
also implies the existence of a scale much smaller than the Planck
mass ($\sim$ 100 TeV) where something other than field theory has
to take over.

It seems possible that a restricted class of such theories may
however be made renormalizable by basing their continuum limit
around a non-perturbative fixed point (rather than the
perturbative Gaussian fixed point that supports the Standard
Model).

Although it has been recognized since the 1970s \cite{Peskin} that
non-Abelian Yang-Mills theory  might have such extra dimensional
ultraviolet fixed points, only a few studies have been made to
search for these. From the lattice studies it seems clear that the
simplest Wilson plaquette bare action does not allow these fixed
points for $SU(2)$ Yang-Mills in $D=5$ or 6 dimensions. However,
as we emphasised, there is no reason to expect the simplest action
to be the correct bare action in this case.\footnote{Actions with
more derivatives than \eq{bulk} would generically have problems
with locality at energies of the overall cutoff, but this cutoff
can be taken to infinity.} The lattice study in ref.
\cite{Nishimura} suggests that even with some more general actions
the fixed points do not exist in $D=6$ dimensions and large $N$.
On the other hand, the exact renormalization group study by Gies
\cite{Gies} suggests that these fixed points do exist in $D=5$
dimensions for $SU(N)$ Yang-Mills at least for $N\le5$.

Somewhat surprisingly, no-one seems to have carried the initial
Wilson epsilon expansion investigation beyond the early two-loop
computation of Peskin \cite{Peskin}, so in this paper we do that
by extending the investigation to the four loops now available. We
find that the $\epsilon$ expansions are very well behaved
asymptotic series, for example we predict that the coefficients in
the large $N$ limit of the expansion for the fixed point coupling
$\ta_*$ do not start to diverge until $\sim$ 10 loops. We give
values for both the fixed point coupling and the critical exponent
$\nu$, together with estimates of the error, by following --where
justified-- the simplest methods possible. In broad terms, it
seems clear that the $D=5$ dimensional fixed points do exist, for
all $N\ge2$. The evidence for fixed points in $D=6$ dimensions, is
marginal, while we find strong evidence that the fixed points do
not exist in any dimension $D\ge7$. As we sought to emphasise, on
the one hand $\epsilon$ expansions above the critical dimension
such as this, are justified from studies in other models, and on
the other hand there is considerable room for improvement on the
present study by using more sophisticated methods.

More generally, within an expansion in $\epsilon$ of the
$D=4+\epsilon$ dimensional theory, these ultraviolet fixed points
exist if and only if the four dimensional theory has only
asymptotically free couplings. By using a dimensional
regularisation and a minimal subtraction type scheme, we can
renormalize the theory perturbatively in the normal way in four
dimensions. Where we need to investigate renormalization group
properties in the higher dimensions, we can use equations such as
\eq{beta} to analytically continue the results to $\epsilon>0$.
These observations open the door to constructing renormalizable
extra dimensional models.

In \sec{MBC}, we extended this generalisation of renormalizability
to include matter and branes. We can add fermions to the bulk
providing there are not too many to destroy the $D=4$ dimensional
asymptotic freedom. It is possible to include bulk scalar fields
but only for careful choices of couplings. The inclusion of both
bulk fermions and scalars is only natural in supersymmetric
theories (where such fixed points have been independently
discovered \cite{Seiberg}). Abelian gauge fields cannot however be
added to the bulk.

Several ideas in the literature such as Higgsless theories, or the
inclusion of Wilson lines directly in the bare action to generate
Yukawa terms, do not extend to renormalizable theories in this
way, at least without further development (see secs. \ref{MBC} and
\ref{GHU}).

When including branes, care has to be taken to work in the correct
critical dimension for the divergences being studied. Thus in the
case of $D=5$ dimensions compactified on an orbifold $S^1/Z_2$,
brane kinetic terms are required from renormalization in $D=5$
dimensions \cite{Georgi} (and not $D=4$ analytically continued to
5). Furthermore, we showed that even if these bulk-generated
divergences were eliminated in a particular model, brane-bulk
interactions would lead to a non-renormalizable theory if the
brane kinetic terms were set to zero. The fact that these are
avoided for non-zero brane kinetic terms is a gauge theory
analogue of the DGP effect \cite{0005016}, and incidentally
implies that the beta functions of the brane-localised fields must
diverge in the limit that the brane kinetic terms are turned off.
These observations have relevance for all such extra dimensional
models, not only the renormalizable ones being proposed here.

We further pursue the phenomenological and theoretical constraints
placed on renormalizable extra dimensional models in \sec{GHU}. We
concentrate on the weak interactions and gauge-Higgs unification
via the Hosotani mechanism \cite{Hosotani}: if such models can be
made renormalizable, they become solutions to the hierarchy
problem. We focus only on the simplest model of gauge-Higgs
unification based on $D=5$ dimensional $SU(3)$ Yang-Mills theory
compactified on $S^1/Z_2$. For the model to be phenomenologically
acceptable, and perturbatively renormalizable in the manner we
have described, we must have large values of $1/b^2\sim10$ for the
$U(1)_Y$ brane kinetic terms, and a much smaller Hosotani vacuum
angle $\theta$ than we would find without some special mechanism.
Indeed we can choose a small $\ta>0.02$ (implying that the effects
of the fixed point will be felt only at energies 10 times higher
than the compactification scale, \ie at energies $\sim 20$ -
50TeV) and choose natural values for the other brane kinetic
terms, in which case we have to arrange the model to dynamically
determine $\theta/2\pi\sim0.04-0.09$. Such values would however
ensure that at low energies the distortions of the Standard Model
are sufficiently small not to come into conflict with precision
measurements.

The known fermions must live only on the brane and do not make a
contribution to the bulk $\beta$ functions (at least at one loop).
The real problem arises when we consider how to generate effective
Yukawa couplings. We seem to be forced to add bulk fermions in
representations and number that are too large to maintain
asymptotic freedom in $D=4$ dimensions, thus destroying the
non-trivial ultraviolet fixed point in $4+\epsilon$ dimensions.
Note that there is no direct relation between this consequence and
the fact that in four dimensions one cannot maintain asymptotic
freedom and add sufficiently many scalars to spontaneously break
all directions in the gauge group \cite{ColemanGross}: the group
in higher dimensions can be much larger and broken in the first
place by the orbifold boundary conditions.

Indeed, in order to make progress, one should consider a larger
group and more involved compactifications. As we noted in
\sec{GHU}, the Standard Model fermions can be encouraged to share
their interactions with the bulk fermions, again reducing the
problem, and leading to a new type of unification, the
possibilities and consequences of which deserve further
exploration.

\bigskip
\acknowledgments

It is a pleasure to thank the following people for useful
discussions: Luis Alvarez-Gaum\'e, Laura Covi, Gian Guidice, Tim
Jones, Riccardo Rattazzi, Douglas Ross, Kari Rummukainen and
Claudio Scrucca.

\end{document}